\documentclass[12pt]{amsart}
\usepackage{graphicx}
\usepackage{amsmath,amsthm}
\usepackage{amssymb}
\usepackage{caption}
\usepackage[latin1]{inputenc}
\usepackage{enumerate}
\usepackage{datetime}

\usepackage[margin=2cm]{geometry} 
\usepackage{relsize}

\usepackage[all]{xy}

\theoremstyle{definition}

\theoremstyle{plain}

\newtheorem*{thm*}{Teorema}
\newtheorem*{cor*}{Corolario}
\newtheorem*{lema*}{Lema}
\newtheorem*{prop*}{Proposición}
\theoremstyle{definition}
\newtheorem*{defi*}{Definición}
\newtheorem*{Ej*}{Ejemplo}
\newtheorem*{obs*}{Nota}
\newtheorem*{Obs*}{Notas}

\numberwithin{equation}{section}

\def\be{\begin{equation}
}

\def\ee{\end{equation}
}

\def\R{\mathbb{R}}

\renewenvironment{abstract}
 {\small
  \begin{center}
  \bfseries \abstractname\vspace{-.5em}\vspace{0pt}
  \end{center}
  \vskip .3cm
  \list{}{
    \setlength{\leftmargin}{.5cm}%
    \setlength{\rightmargin}{\leftmargin}%
  }%
  \item\relax}
 {\endlist}

\begin{document}

\title[Redshift for massive particles]
 {Redshift for massive particles}

\author{J.  Mu\~{n}oz-D{\'\i}az and R. J.  Alonso-Blanco}

\address{Departamento de Matem\'{a}ticas, Universidad de Salamanca, Plaza de la Merced 1-4, E-37008 Salamanca,  Spain.}
\email{ricardo@usal.es, clint@usal.es}

\maketitle


\begin{abstract}
On each FRW metric, there exist a redshift for massive particles, analogous to that of photons, but depending on the mass. This is a simple consequence of the fundamental laws of Classical Mechanics.
\end{abstract}
\vskip 1cm

Let us consider a manifold $M=\R\times\overline M$ endowed with a metric
$$T_2=dz^2-a(z)\,\overline T_2$$
where $z$ is the ``cosmological time'' and $\overline T_2$ is a Riemannian ``spatial metric'' on $\overline M$.

The Liouville form on $TM$ associated with $T_2$ is
$$\theta=\dot z\,dz-a(z)\,\overline\theta$$
where $\overline\theta$ is the Liouville form on $\overline M$ associated with $\overline T_2$.

The function on $TM$ associated with  the horizontal 1-form $\theta$ (see \cite{MecanicaMunoz}) is
$$\dot\theta=\dot z^2-a(z)\,\dot{\overline\theta}$$
which is 2 times the ``kinetic energy'' $T$.

Let $D$ be the geodesic field on $TM$ associated with the metric $T_2$. The mechanical system $(M,D)$ is conservative with Lagrangian
$$L=T=\frac 12\,\dot\theta.$$

In general, each tangent vector field $u$ on a manifold $M$ determines a function on $T^*M$, defined by 
$$p_u:=\langle\theta,u\rangle$$
(inner contraction), 
where $\theta$ is the Liouville form moved to $T^*M$ (by means of the metric). The function $p_u$ is the \emph{momentum associated with $u$}. When $M$ is the configuration space of a conservative mechanical system $(M,T_2,dU)$, with Lagrangian $L=T-U$, the evolution of the momentum function $p_u$ is given by the Hamilton-Noether Equation (see \cite{MecanicaMunoz}, p.80, \cite{RM}):
$$Dp_u=\delta_uL,$$
where $\delta_u$ is the infinitesimal variation defined by $u$ (see \cite{MecanicaMunoz}).

In our case, we take $u=\textrm{grad}\, z=\partial/\partial z$ (so that $\delta_u$ is also $\partial/\partial z$) and denote the momentum associated by $p_0$:
$$p_0=\langle\theta,\partial/\partial z\rangle=\dot z.$$

The Hamilton-Noether Equation along each geodesic is 
$$\frac{dp_0}{dt}(=Dp_0)=\frac{\partial}{\partial z}\left(\frac 12\,\dot\theta\right)=-\frac 12 a'(z)\,\dot{\overline\theta}.$$
Here, $t$ is the parameter of the geodesic, which in our case equals the proper time (see \cite{RelatividadMunoz}).

Multiplying by $$dt/dz=\dot z^{-1}=p_0^{-1}$$ we obtain from the above,
$$\frac{dp_0}{dz}=-\frac 12 a'(z)\,\frac{\dot{\overline\theta}}{p_0}.$$

For light trajectories is $\dot\theta=0$, so that, $\dot{\overline\theta}=\dot z^2/a(z)=p_0^2/a(z)$, and we get
$$\frac{dp_0}{dz}=-\frac 12 \frac{a'(z)}{a(z)}\,{p_0},$$
or,
$$\frac{dp_0}{p_0}+\frac 12\frac{da}a=0,$$
which gives
$$p_0=\textrm{const.}\,a(z)^{-1/2}.$$

Because the ``radius of the Universe'' $R$ is $\varpropto a(z)^{1/2}$ we have
$$p_0\varpropto R(z)^{-1},$$
which is the classical law for the redshift of the light.
\bigskip

Let us consider now a geodesic into the  mass-shell $\dot\theta=\mu$, $\mu>0$. The Hamilton-Noether Equation (or, alternatively, the Geodesics Equations) gives, along each geodesic
$$\frac{dp_0}{dz}=-\frac 12 a'(z)\,\frac{\dot{\overline\theta}}{p_0}=-\frac 12 a'(z)\,\frac{1}{p_0}\left(\frac{p_0^2-\mu}{a(z)}\right).$$

The general solution of the above differential equation is
$$(p_0^2-\mu)\,a(z)=\textrm{const.}$$

By substituting $\mu$ by its value on a given geodesic,
$\mu=\dot\theta=p_0^2-a(z)\dot{\overline\theta}$, we get $\dot{\overline\theta}\,a(z)^2=\textrm{const.}$, or,
\begin{equation}\label{redshift}
\dot{\overline\theta}=\textrm{const.}\,a(z)^{-2}.
\end{equation}

That is the law of redshift for the ``kinetic energy'' on the spatial manifold  $\overline M$ of a particle  moving along a geodesic on $M$.

For a light trajectory is 
$$p_0^2=a(z)\,\dot{\overline\theta},$$
and the general rule (\ref{redshift}) gives again 
$$p_0\varpropto a(z)^{-1/2}.$$

\vskip 1cm


\end{document}